\documentclass[aps,prb,reprint,amsmath,amssymb,superscriptaddress,longbibliography]{revtex4-2}
\usepackage{color,graphicx,float,hyperref,stackengine,braket,bm,mathtools,calc}
\usepackage[caption=false]{subfig}
\def\dvec#1{\overset{\text{\tiny$\leftrightarrow$}}{#1}}

\begin{document}

\title{Electron ground state $g$ factor in embedded InGaAs quantum dots: \\ An atomistic study}

\author{Mustafa  \surname{Kahraman}}
\author{Ceyhun \surname{Bulutay}}
\email{bulutay@fen.bilkent.edu.tr}
\affiliation{Department of Physics, Bilkent University, Ankara 06800, Turkey}

\date{\today}

\begin{abstract}
We present atomistic computations within an empirical pseudopotential framework for the
electron $s$-shell ground state $g$ tensor of InGaAs quantum dots (QDs) embedded to host matrices that 
grant electronic confinement.
A large structural set consisting of geometry, size, and molar fraction variations is worked out
which also includes a few representative uniform strain cases. The tensor components are observed to 
display insignificant discrepancies even for the highly anisotropic shapes. The family of $g$-factor 
curves associated with these parameter combinations coalesces to a single universal one when plotted 
as a function of the gap energy, thus confirming a recent assertion reached under much 
restrictive conditions. Our work extends its validity to alloy QDs with various shapes 
and finite confinement that allows for penetration to the host matrix, placing it on a more
realistic basis. Accordingly, the electrons in InGaAs QDs having $s$-shell transition energies 
close to 1.13~eV will be least susceptible to magnetic field. We also show that low indium 
concentration offer limited $g$-factor tunability under shape or confinement 
variations. These findings can be taken into consideration in the fabrication 
and the use of InGaAs QDs with $g$-near-zero or other targeted $g$ values for spintronic or electron spin 
resonance-based direct quantum logic applications.
 
\end{abstract}

\maketitle

\section{Introduction}

A single parameter, namely the $g$ factor, succinctly provides a measure of how strongly a charge in an 
electronic structure couples to an external magnetic field. Compared to its free-electron Dirac 
equation value of $g_0=2$, it can be significantly renormalized in solids, denoted by $g^*$, as a 
result of the spin-orbit interaction \cite{yafet63,callaway}. Likewise, in semiconductor 
nanostructures such as quantum dots (QDs), yet another level of renormalization becomes operational 
by confining the carrier wave function around a heterogeneous region which accordingly tailors 
the orbital contribution \cite{sandoval18}, at the same time offering electrical tunability 
\cite{nakaoka07,kato08,klotz10,pingenot11,taylor13}. Among these structures, the self-assembled 
InGaAs QDs particularly stand out where a number of critical quantum technological milestones 
have been demonstrated, like indistinguishable single-photon sources \cite{santori02}, also 
on demand \cite{he13}, spin-resolved resonance fluorescence \cite{vamivakas09}, spin-photon interface 
\cite{yilmaz10}, entangled photon pairs \cite{stevenson08}, entanglement swapping \cite{gao12}, 
as well as simultaneous antibunching and squeezing \cite{schulte15}. Interestingly, the electron 
spin resonance (ESR) in \textit{embedded} InGaAs QDs has so far been elusive, with the exception of one 
report which awaits to be reproduced for more than a decade \cite{kroner08}. As a matter of fact, 
ESR would be highly welcomed in embedded QDs for the direct magnetic field control of the electron 
spin over the full Bloch sphere, which was unambiguously showcased much earlier in electrostatically 
defined gated QDs \cite{koppens06}.

An intriguing region that similarly merits attention is where the QD $g$-factor changes sign, which is
of relevance to $g$-near-zero QDs ($g^*\sim 0$). There are a number of reasons why this can bring 
interesting physics. In general the background nuclear spins interact with external magnetic field with a 
coupling constant about three orders of magnitude smaller than those of free electrons, originating 
from their Land\'{e} factor ratio \cite{urbaszek13}. Consequently, nuclear magnetic resonance (NMR) and ESR 
frequencies are off by again a factor of thousand. $g$-near-zero QDs mitigate the ESR and 
NMR mismatch so that the electron-nucleus counter spin flips become energetically more affordable. 
This can be utilized to achieve a strong coupling between electron and the nuclear spin bath, similar 
to the Hartmann-Hahn double resonance \cite{hartmann62,london13,liu14}. 
From a basic science point of view, $g$-near-zero can promote a spin-density wave state
where the spins are oriented perpendicular to the magnetic field \cite{julian11}, and a spin texture of skyrmionic 
excitations \cite{mitrovic07}. As to some practical examples, it can facilitate 
controlled spin rotation by $g^*$-tunability provided that it changes sign via electric gating 
\cite{nakaoka07,ulhaq16}, or the quantum state transfer between a flying photon qubit and a resident 
electron spin qubit in a QD \cite{kuwahara10}. Thus, a deeper understanding of the elements that govern 
the $g$ factor, especially in InGaAs QDs, is quite valuable for several research directions.

Over the years there has been a number of experimental efforts to better characterize the $g$ factor 
of InGaAs QDs 
\cite{bayer99,nakaoka04,nakaoka05,alegre06,schwan11a,schwan11b,sapienza16,tholen16,tholen19,wu20}. 
A complication that arises in most of these magnetoluminescence-based measurements is the extraction 
from the \textit{excitonic} $g$-factor that of the electronic contribution 
\cite{nakaoka04,nakaoka05,klotz10,schwan11a,schwan11b,taylor13,tholen16,tholen19}, 
which hinders its sign, a concern also shared by the magnetocapacitance \cite{alegre06}, and 
photocurrent spectroscopy experiments \cite{wu20}. Naturally, they need to be supplemented by an 
electronic structure theory, which has been routinely a variant of the $k\cdot p$ model 
\cite{kiselev98,pryor06,zielke14,gawarecki18,mielnik18}, even though more sophisticated alternatives 
are being developed \cite{delerue-book,usman12a,pryor15,mittelstadt19}. 
Another difficulty that virtually affects all experimental studies stems from not knowing the precise structural 
information such as the alloy composition, geometry and hence the strain profile of the \textit{probed} single 
QD. Significant progress was put forward by an inverse approach by feeding in spectroscopical data into theory,
to find structural models that agree with both the cross-sectional scanning tunneling microscopy and spectroscopy 
measurements \cite{mlinar09a,mlinar09b}. However, an ambiguity still remains, as such an approach might overlook the 
subtleties due to the decaying indium concentration above the QD and the extension of the QD into the 
wetting layer \cite{giddings11,keizer11}.
Notwithstanding, as reported by some of these works, this deficient knowledge may not be so critical as
it is primarily the gap energy that is directly linked with the $g$ factor \cite{kato08,schwan11a}, 
by that they substantiate the celebrated so-called Roth-Lax-Zwerdling expression which was originally 
derived for bulk \cite{roth59}. A recent tight-binding analysis qualitatively supported this by 
concluding that the dominant contribution to the $g$ factor of nanostructures comes from the 
bulk term, considering only the compound QDs \cite{tadjine17}. Undoubtedly, these assertions merit further 
theoretical investigation, preferably by an atomistic electronic structure technique that can grant 
quantitative insights into \textit{alloy} InGaAs QDs as it is predominantly for the ones studied so far  
\cite{bayer99,nakaoka04,nakaoka05,nakaoka07,klotz10,pingenot11,schwan11a,schwan11b}.

In this work, we consider In$_x$Ga$_{1-x}$As QDs under a homogeneous compressive hydrostatic strain 
characteristic of the inner cores of the partially relaxed structures \cite{guffarth01,usman12b,bulutay12}. 
As the embedding material, GaAs is by far the most common choice, but its confinement is rather limited 
by its bulk band gap of 1.52~eV. Other compatible wider gap options are available, such as In$_x$Al$_{1-x}$As 
with a room temperature energy gap above 2~eV \cite{golovynskyi21}, and in the case of 
(In$_x$Ga$_{1-x}$)$_2$O$_3$ this reaches 5~eV \cite{wenckstern15}.
Therefore, we assume that each studied QD is embedded into a sufficiently wide band gap matrix 
that provides confinement for the $s$-shell ground state electron. 
The QD geometries worked out range from full spherical up to a lens shape as well as various cuts in 
between with respect to the [111] axis. Independently, the indium fraction is also varied within a uniform 
alloy profile inside the QD. These structures embody on the order of 10 million atoms including 
the matrix material which makes it imperative to use an efficient atomistic electronic structure tool. 
In our case we employ the so-called linear combination of bulk bands (LCBB) which handles 
such atomic numbers with reasonable computational budget \cite{wang99}. In the past, we used it 
in nanocrystals for the linear optical response \cite{bulutay07}, third-order nonlinear optics 
\cite{yildirim08}, electroabsorption \cite{bulutay10}, and coherent population transfer \cite{gunceler10},
and in nanowire structures for electronic structure \cite{keles13} and ballistic transport \cite{keles15}. 

Most importantly, among other findings, our work substantiates the conclusion of the aforementioned 
tight-binding study which reported a universal behavior for $g^*$ when plotted with respect to the 
gap energy \cite{tadjine17}. Furthermore, our fitted $g$-factor curve applies to alloy 
QDs of various shapes with finite confinement that allows for penetration to the matrix as in realistic 
samples. This result can be beneficial in InGaAs QDs for both achieving a well-controlled 
ESR as well as avoiding it, depending on the specific purpose. As for the former, if a successful and 
reproducible ESR is aimed, in the very unfavorable signal-to-noise ratio due to vibrant nuclear spin 
background \cite{kroner08},
there should be no room for ambiguity in the precise frequency of the ESR, hence the $g$ factor. For that matter 
our fit enables a simple estimate for it based solely on the transition energy, without knowing
the precise molar composition and structural information of the QD. Going to the other extreme, if the
magnetic effects are desired to be minimized for the $s$-shell ground state electron, as for instance to 
protect the electron spin qubit \cite{bechtold15}, then InGaAs QDs with transition energies close to 1.13~eV 
should be targeted according to our analysis that predicts for them $g^*\sim 0$.

The organization of the paper is as follows: In Sec.~II we describe the LCBB technique as it is 
not widely known, together with the $g$-factor expressions. Our computational implementation 
determines the constraints under which we perform the calculations; in this respect they are crucial, 
and included in Sec.~III. Section~IV presents our results for a rich variety of QD structures, 
and reveals the underlying universal behavior, followed by our conclusions in Sec.~V. 

\section{Theory}
\subsection{LCBB Electronic Structure Technique}
A general necessity in atomistic electronic structure techniques is a large basis set,
as in the form of extended plane waves or localized Gaussian orbitals,
so as not to compromise accuracy, which inevitably inflates the computational budget. 
Yet, when a restricted energy window is of interest, a specialized basis set of modest 
size selected with physical insight becomes viable, forming the premise of the 
LCBB method \cite{wang99}. Here, the basis set is formed by the bulk Bloch 
functions of the underlying materials within the desired energy range. 
Hence, the $j^{th}$ stationary state wave function of a nanostructure is approximated by the expansion
\begin{equation}
\psi_{j}(\mathbf{r})
=
\frac{1}{\sqrt{N}}
\sum_{n,\mathbf{k},\mu}
C^{\mu,j}_{n\mathbf{k}}
u^{\mu}_{n\mathbf{k}}(\mathbf{r}) e^{i\mathbf{k}\cdot\mathbf{r}} ,
\end{equation}
where $N$ is the number of primitive unit cells inside the
large supercell of the nanostructure, $n$ is the bulk band index,
$\mathbf{k}$ is the wave vector within the first Brillouin zone of the
underlying lattice, and $\mu$ designates the materials in the set,
usually the core and the embedding media. In this expression the cell-periodic part 
$u^{\mu}_{n\mathbf{k}}(\mathbf{r})$ of the bulk
Bloch functions of each material has the Fourier series representation
\begin{equation*}
 u^{\mu}_{n\mathbf{k}}(\mathbf{r})=\frac{1}{\sqrt{\Omega_0}}\sum_\mathbf{G}
 B^{\mu}_{n\mathbf{k}}(\mathbf{G})e^{i\mathbf{G}\cdot\mathbf{r}}\, ,
\end{equation*}
where the summation is over the reciprocal lattice vectors $\mathbf{G}$, 
inside an energy cut-off, and $\Omega_0$ is the volume of the primitive cell \cite{callaway}. 
The Fourier coefficients $B^{\mu}_{n\mathbf{k}}(\mathbf{G})$ are readily accessible 
by diagonalizing the bulk Hamiltonian of material $\mu$ at each $\mathbf{k}$ point.

The single-particle Hamiltonian of a nanostructure constitutes the kinetic energy 
and the crystal potential parts. For the latter we employ the empirical pseudopotentials 
\cite{bester09} to describe the atomistic environment
\begin{eqnarray*} \label{Hamiltonian}
\mathcal{H} & = & \mathcal{T} + \mathcal{V}_\mathrm{xtal} \\
 & = & -\frac{\hbar^2\nabla^2}{2m_0}+
\sum_{\mu,\mathbf{R}_l,\alpha} W^{\mu}_{\alpha}(\mathbf{R}_l)\,
\upsilon^{\mu}_{\alpha} ( \mathbf{r}-\mathbf{R}_l-\mathbf{d}^{\mu}_{\alpha}) \, ,
\end{eqnarray*}
where $m_0$ is the free electron mass, the direct lattice vector
$\mathbf{R}_l$ indicates the origin for each primitive cell, $l$, and
$\mathbf{d}^{\mu}_{\alpha}$ specifies the relative coordinate of the basis atom 
$\alpha$ within the primitive cell. The weight function
$W^{\mu}_{\alpha}(\mathbf{R}_l)$ keeps the information about the atomistic
composition of the nanostructure by taking values 0 or 1 depending on the
type of the atom located at the position
$\mathbf{R}_l+\mathbf{d}^{\mu}_{\alpha}$.
$\upsilon^{\mu}_{\alpha}$ is the local screened spherical
atomic pseudopotential of atom $\alpha$ of the material $\mu$ \cite{bester09}.

Hamiltonian matrix elements are evaluated with respect to the LCBB basis states,
\{$\vert n\mathbf{k}\mu \rangle$\} which can be cast into a generalized eigenvalue problem
\begin{equation*}\label{sch} 
\sum_{n,\mathbf{k},\mu} \hspace{-3pt}  \langle n^{\prime}\mathbf{k}^{\prime}
\mu^{\prime}\vert\mathcal{T}+\mathcal{V}_\mathrm{xtal}\vert n\mathbf{k} \mu
\rangle \, C^{\mu,j}_{n\mathbf{k}} \hspace{-2pt} = \hspace{-3pt} E_j \hspace{-3pt} \sum_{n,\mathbf{k},\mu} \hspace{-3pt}
C^{\mu,j}_{n\mathbf{k}} \langle
n^{\prime}\mathbf{k}^{\prime}\mu^{\prime}\vert n\mathbf{k}\mu
\rangle  \, ,
\end{equation*}
which yields the energy $E_j$ and the expansion coefficients $C^{\mu,j}_{n\mathbf{k}}$.
The explicit forms of these matrix elements are
\begin{equation*}\label{bulk11}
\langle n^{\prime}\mathbf{k}^{\prime}\mu^{\prime}\vert
n\mathbf{k}\mu \rangle = \delta_{\mathbf{k},\mathbf{k}^{\prime}}
\sum_{\mathbf{G}} 
\big[B^{\mu^{\prime}}_{n^{\prime}\mathbf{k}}(\mathbf{G})\big]^*
B^{\mu}_{n\mathbf{k}}(\mathbf{G}) \, ,
\end{equation*}
\begin{equation*}\label{Tmat1}
\langle n^{\prime}\mathbf{k}^{\prime}\mu^{\prime} \vert \mathcal{T}
\vert n\mathbf{k}\mu \rangle = \delta_{\mathbf{k},\mathbf{k}^{\prime}}
\sum_{\mathbf{G}} \frac{\hbar^2 \vert \mathbf{k}+\mathbf{G} \vert^2}{2m_0}
\big[B^{\mu^{\prime}}_{n^{\prime}\mathbf{k}}(\mathbf{G})\big]^*
B^{\mu}_{n\mathbf{k}}(\mathbf{G}) \, ,
\end{equation*}
\begin{eqnarray*}
\langle n^{\prime}\mathbf{k}^{\prime}\mu^{\prime} \vert
\mathcal{V}_\mathrm{xtal} \vert n\mathbf{k}\mu \rangle =
& \sum_{\mathbf{G},\mathbf{G}^{\prime}} 
\big[B^{\mu^\prime}_{n^\prime\mathbf{k}^\prime}(\mathbf{G}^\prime)\big]^*
B^{\mu}_{n\mathbf{k}}(\mathbf{G}) \\ & \times
\sum_{{\mu^{\prime\prime}},\alpha}
\mathcal{V}_\alpha^{{\mu^{\prime\prime}}} (\vert
\mathbf{k}+\mathbf{G}-\mathbf{k}^{\prime} -\mathbf{G}^{\prime} \vert)
 \\ & \times  \mathcal{W}_\alpha^{{\mu^{\prime\prime}}}(\mathbf{k}-\mathbf{k}^{\prime})
e^{-i(\mathbf{k}+\mathbf{G}-\mathbf{k}^{\prime} -\mathbf{G}^{\prime})
\cdot\mathbf{d}_\alpha^{{\mu^{\prime\prime}}}} \, .
\end{eqnarray*}
Here, $\mathcal{V}_\alpha^{{\mu^{\prime\prime}}}$ and $\mathcal{W}_\alpha^{{\mu^{\prime\prime}}}$
are the Fourier transforms of atomic pseudopotentials and the weight functions
\begin{eqnarray}
\mathcal{V}_\alpha^{{\mu^{\prime\prime}}} (\vert \mathbf{k}+\mathbf{G}-\mathbf{k}^{\prime} 
-\mathbf{G}^{\prime} \vert)
& = & 
\frac{1}{\Omega_0} \int\upsilon_{\alpha}^{\mu^{\prime \prime}}(\mathbf{r})
e^{i(\mathbf{k}+\mathbf{G}-\mathbf{k}^{\prime}-\mathbf{G}^{\prime})\cdot \mathbf{r}} d^3r, 
\label{VFT} \nonumber \\
 & & \\
\mathcal{W}_{\alpha}^{\mu}(\mathbf{k}-\mathbf{k}^{\prime})
& = &
\sum_{j} W_{\alpha}^{\mu^{\prime {\prime}}}(\mathbf{R}_j) e^{i(\mathbf{k}-\mathbf{k}^{\prime})\cdot \mathbf{R}_j} .
\label{DFT}
\end{eqnarray}

\subsection{Spin-orbit interaction}
So far, only the spin independent part of the Hamiltonian is considered. 
Following Hybertsen and Louie \cite{hybertsen86}, the spin-orbit interaction can be incorporated as 	
\begin{equation}
\mathcal{H}_{SO} = \sum_{\ell=1}^{\infty} \ket{\ell} V^{SO}_\ell(r) \bm{\ell}\cdot \bm{\sigma} 
\bra{\ell} \ ,
\label{SOI}
\end{equation}
where, $\ell$ is the orbital angular momentum label, $\bm{\sigma}$ is the vector Pauli spin operator, and 
$V^{SO}_\ell(r)$ is the angular-momentum-dependent (i.e., nonlocal) radial spin-orbit potential. 
To simplify, we restrict to the dominant $\ell=1$, i.e., $p$ component, and the spin-orbit matrix 
elements become
\begin{eqnarray}
\bra{s,\mathbf{K}}\mathcal{H}_{SO}\ket{s^{\prime},\mathbf{K}^{\prime}} & = & -i \bra{s}\bm{\sigma}
\ket{s^{\prime}} \cdot \bigg[ 12\pi \frac{\mathbf{K}\times \mathbf{K}^{\prime}}
{K K^{\prime}} \nonumber \\  
& & \times V_{\ell=1}^{SO}(K,K^\prime) \bigg] S(\mathbf{K}^{\prime} - \mathbf{K})\ , 
\label{SOI-matrix}
\end{eqnarray}
where $\mathbf{K}=\mathbf{k}+\mathbf{G}$, $\mathbf{K}^\prime=\mathbf{k}+\mathbf{G}^\prime$, 
$\ket{s}$ denotes a spinor state, $S(\mathbf{K}^{\prime} - \mathbf{K})$ 
is the bulk static structure factor. $V_{\ell}^{SO}(K,K^\prime)$ is given by the integral
\begin{equation}
V_{\ell}^{SO}(K,K^\prime) = \int_{0}^{\infty} \frac{dr}{\Omega_0} r^2 j_{\ell}(Kr)\ 
V_{\ell}^{SO}(r)\ j_{\ell}(K^{\prime}r) \ ,
\end{equation}
with $j_{\ell}$ being the spherical Bessel function of the first kind and $V_{\ell}^{SO}(r)$ 
is chosen as a Gaussian function \cite{williamson00} with a width of 2.25 Bohr radius and its amplitude 
being a fit parameter, $\lambda_S$ as described below. $V_{\ell}^{SO}(K,K^\prime)$ is computed once, 
and invoked from a look-up table.

\subsection{$g$ factor}
Unlike a free electron, the charge of a nanostructure experiences an anisotropic coupling to an external
magnetic field $\mathbf{B}$ so that its $g$ factor becomes a rank-2 tensor $\dvec{g}$,
which in the most general case is characterized by nine linearly independent components \cite{kiselev98}. 
It is described through the Zeeman Hamiltonian
\begin{equation}
\mathcal{H}_Z=\frac{1}{2}\mu_B\, \bm{\sigma}\cdot\dvec{g}\cdot\mathbf{B} ,
\end{equation} 
where $\mu_B$ is the Bohr magneton.
The celebrated $\dvec{g}$ expression follows from a spinless electronic structure calculation when the spin-orbit 
interaction is included as a first-order perturbation \cite{roth60,callaway}. 
This bulk formulation can be extended to QDs in terms of the matrix elements between two confined states $n$ and $j$ as
\begin{eqnarray}
	\label{pnj}
	\mathbf{p}_{nj} & = & \frac{(2\pi)^3}{\Omega_\mathrm{SC}}\int_\mathrm{SC} \psi_n^*(\mathbf{r})\ \mathbf{p}\ \psi_j(\mathbf{r})\ d^3r \, , \\
	\label{hnj}
	\mathbf{h}_{nj} & = & \frac{(2\pi)^3}{\Omega_\mathrm{SC}}\int_\mathrm{SC} \psi_n^*(\mathbf{r})\ \mathbf{h}\ \psi_j(\mathbf{r})\ d^3r \, ,
\end{eqnarray}
where the integrals are over the supercell volume $\Omega_\mathrm{SC}$, $\mathbf{p}$ 
is the momentum operator, and $\mathbf{h}$ is the spin-orbit related operator defined through 
$\mathcal{H}_{SO} = \mathbf{h}\cdot\bm{\sigma}$; see, Eq.~(\ref{SOI}). 

For a chosen state $n$, this yields the $g$ factor
\begin{eqnarray}\label{gn}
	\dvec{g}_n  & = & 2\dvec{I} + \frac{2}{i \hbar^2 m_0} 
	\sum_{jl}{\vphantom{\sum}}' \frac{1}{\omega_{nj}} \bigg[ \frac{(\mathbf{h}_{jl}-\mathbf{h}_{lj})
	(\mathbf{p}_{nj}\times\mathbf{p}_{ln})}{\omega_{jl}} \nonumber\\
	& & +\frac{(\mathbf{h}_{ln}-\mathbf{h}_{nl})(\mathbf{p}_{nj}\times\mathbf{p}_{jl})}{\omega_{nl}} \bigg],
\end{eqnarray}
where $\dvec{I}$ is the 3$\times$3 identity matrix, the prime over the summation stands for $j\ne l$, 
and $\omega_{nj}=(E_n-E_j)/\hbar$, and etc. With some manipulations, it can be shown to be equivalent to 
the Roth's bulk expression \cite{roth60}
\begin{eqnarray}\label{roth}
 \dvec{g}_n & = & 2\dvec I + \frac{2}{i\hbar^2 m_0} \sum_{jl}{\vphantom{\sum}}' 
 \frac{1}{\omega_{nj} \omega_{nl}} \left(\mathbf{h}_{nj}\mathbf{p}_{jl}\times \mathbf{p}_{ln} \right.\nonumber\\
 & & + \left. \mathbf{h}_{jl}\mathbf{p}_{nj}\times \mathbf{p}_{ln} + \mathbf{h}_{ln}\mathbf{p}_{nj}\times \mathbf{p}_{jl}\right),
\end{eqnarray}
where in contrast to bulk, here the matrix elements are worked out using nanostructure states 
as given by Eqs.~(\ref{pnj}) and (\ref{hnj}).

\section{Computational implementation}
In this section we would like to give some important details about our computational 
model. Foremost, we utilize our recently fitted empirical pseudopotentials for InAs and GaAs 
under various strain conditions to hybrid density functional theory band structures \cite{cakan16}. 
In anticipation to reduce matrix sizes, the fit was achieved with about 120 reciprocal lattice vectors 
within the energy cut-off. For the current work involving QDs having the In$_x$Ga$_{1-x}$As alloy core, 
we use Vegard's law in mixing the compound InAs and GaAs pseudopotentials. 
As mentioned above, the spin-orbit interaction over the $p$ states brings a further symmetric 
spin-orbit coupling parameter $\lambda_S$ fitted to experimental spin-orbit splittings for 
bulk InAs and GaAs \cite{cakan16}.

The $\dvec{g}_n$ expression in Eq.~(\ref{roth}) requires, in principle, all of the QD states, but especially 
those energetically close to the state $n$ under investigation. This demands well characterization of a large 
number of electronic states which hinges upon the strength of the LCBB basis set. Recall that in our formulation 
the spin-orbit interaction enters as a perturbation, hence the wave functions are spinless.
We employ the bulk bands of the spinless top four (four) valence and the lowest four (one) conduction bands 
of the strained core (matrix) material. For either case, basis sets are formed from a three-dimensional  
$5\times 5\times 5$ grid in the reciprocal space centered around the $\Gamma$ point. 
Its convergence was checked for the supercell size we adopted for calculations in this work. 
The final LCBB basis sets typically contain some two thousand elements.

The non-self-consistent nature of the empirical pseudopotentials \cite{bester09} entails an additional 
bulk parameter to have a desirable band alignment under strain. Following Williamson \emph{et al}, this is implemented 
as a hydrostatic strain-dependent pseudopotential formed as
\begin{equation}
V(q;\epsilon)=\left[ 1+\gamma\,\epsilon_H\right]\, V(q)\, ,
\label{strain-dependent-psp}
\end{equation}
where $\gamma$ is the accompanying fitting parameter and $\epsilon_H=\epsilon_{xx}+\epsilon_{yy}+\epsilon_{zz}$ 
refers to hydrostatic strain \cite{williamson00}. 
We should note that in all of the calculations, we assume a \emph{uniformly} strained QD so that
the same lattice constant applies over the full supercell. This greatly simplifies the computational 
tasks like the choice of the basis sets, and allows the use of the standard fast Fourier transform (FFT) in 
Eq.~(\ref{DFT}) \cite{fftw}. Even then, representing the existing theoretical and experimental 
band offset data \cite{walle89,pryor98,gosh00,usman09} becomes quite challenging mainly due to strain-related 
band gap as well as line-up variations for all the structures worked out in this study. This necessitates the 
use of several different artificial matrices, in each case lattice-matched to core (strained) QD, and with band gaps 
ranging from 1.52~eV (GaAs) to 5~eV [such as (In$_x$Ga$_{1-x}$)$_2$O$_3$]. The conduction band offset values 
depending on indium mole fraction and strain range between 65~meV to 245~meV to ensure the confinement 
$s$-shell ground state electron.

In regard to the above simplifications of our computational model, the pioneering works expressed that 
the embedded InGaAs QD geometry and the structural relaxation result in a position-dependent strain within 
the QD, see for e.g., \cite{pryor98,williamson00,guffarth01}. A number of $k\cdot p$ and tight-binding studies 
elaborated on their electronic properties using highly sophisticated QD structural information, 
see for e.g., \cite{bimberg-book,klimeck07,delerue-book}. On the other hand, from the LCBB point of view 
inhomogeneous strain would call for \emph{nonuniform} FFT which slows down the calculations drastically, 
even taking into account recently 
developed packages \cite{nufft1,nufft2}. Moreover, a much richer strained basis set is required that compounds 
the computational overhead. For this reason, we opt for a few uniform cases, characteristic of average strain 
present in typical QDs  \cite{guffarth01,usman12b,bulutay12}. This leaves outside the effect of inhomogeneous 
strain on the $g$ factor, thereby pinpointing a direction along which our approach can be further improved.

\section{Results}
\subsection{Cuts from a sphere}
We start with the compound InAs spherical QD of 45~nm diameter embedded in a host matrix, where the QD is 
under a 2\% compressive strain, corresponding to a hydrostatic strain of $\epsilon_H=-0.06$. In Fig.~1 
we see how the principal values of $\dvec{g}$ vary when the sphere is successively cut by a (111) 
oriented plane, producing in addition to a sphere, a hydrophobic-contact-angle-, hemispherical-, and 
lens-shaped QD. As expected, the increasing confinement gradually modifies $g^*$ from -2.47 to 0.21 so 
that $g^*\sim 0$ would be attained for a lens shaped QD with a bigger diameter than the one in Fig.~1. 
The three principal values of $\dvec{g}$ marginally deviate from each other even though 
they become exceedingly of anisotropic shapes toward the lens QD. 
The largest difference is about 0.03 that occurs for the spherical QD which indicates the numerical 
accuracy of our calculations. This lack of anisotropy in $\dvec{g}$ is ubiquitous 
for all the structures studied in this work. Therefore, we shall display its major principal 
component in the plots to follow.

\begin{figure}
\centering
\includegraphics[width=1.0\columnwidth]{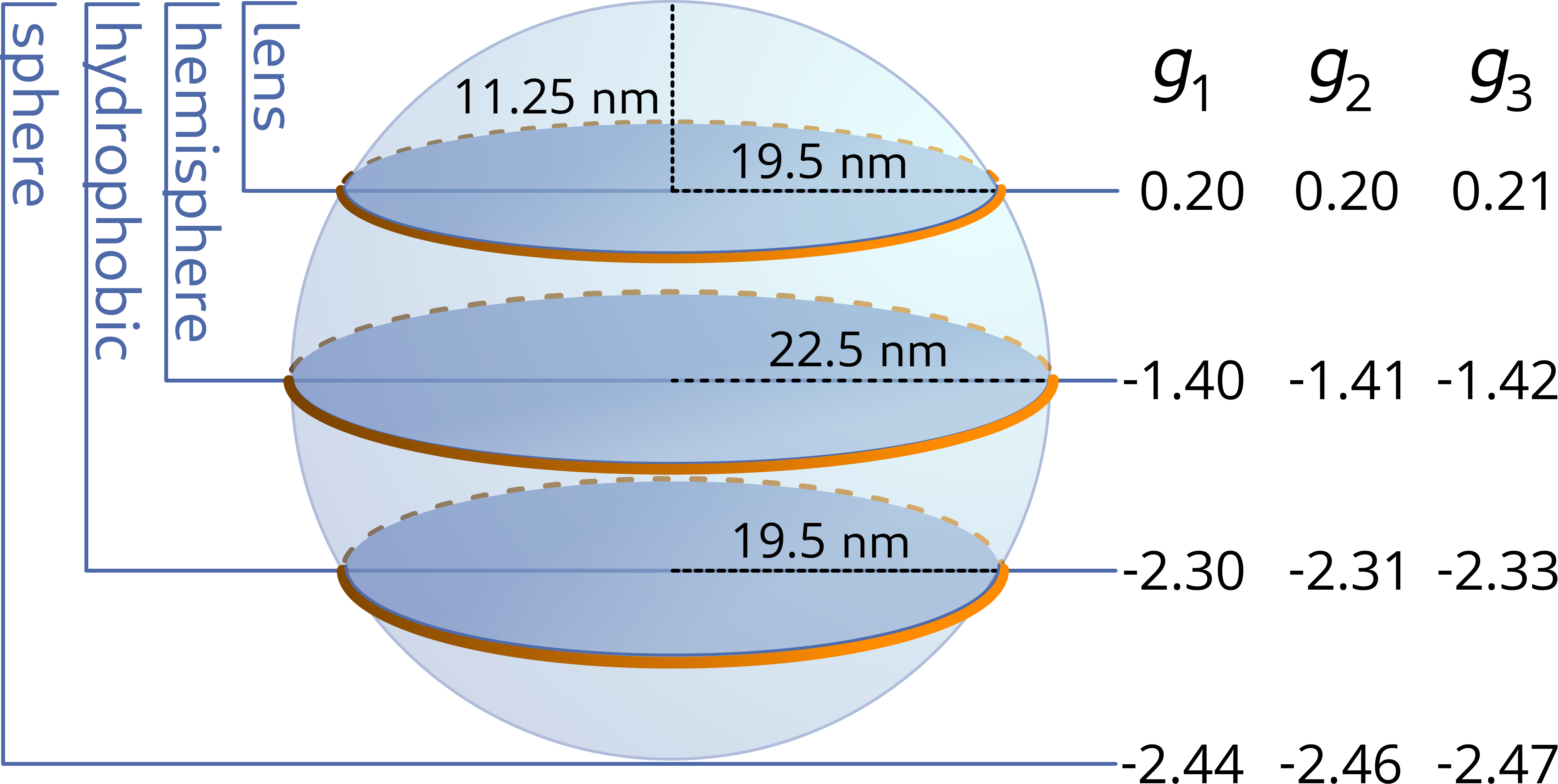}
\caption{The three principal $\dvec{g}$ values of embedded InAs QDs  
under 2\% homogeneous compressive strain. The four geometries originate from a sphere by cutting 
with a (111) plane producing lens, hemisphere, and hydrophobic-contact-angle spherical domes.}
\label{fig1}
\end{figure}

Next, choosing the spherical, hemispherical, and lens geometries from Fig.~1, we consider how both 
$g$ factor and the highest occupied molecular orbital (HOMO) to lowest unoccupied molecular 
orbital (LUMO) energy gap evolve with the indium molar fraction for alloy In$_x$Ga$_{1-x}$As QDs.
Figure~2 illustrates the family of curves belonging to each shape with a diameter around 46~nm, 
and for the lens ones of height about 11-12~nm for the QD strain value of -2\%, i.e., 
$\epsilon_{xx}=\epsilon_{yy}=\epsilon_{zz}=-0.02$. In Fig.~2(a) the geometric sensitivity in $g^*$ 
reveals itself toward the indium-rich composition, where the sign change also takes effect. In that 
respect, indium-poor QDs offer very limited $g$-tunability. 
The accompanying energy gaps in Fig.~2(b) merit some explanation as one would expect an opposite trend 
based on the quantum size effect consideration which would decrease as the indium content is reduced toward 
GaAs because of the heavier effective mass of GaAs. This tendency is more than 
compensated by the increase in the gap energy due to increasing gallium fraction. It is considerably 
boosted under strain as the bulk band gap deformation potential of GaAs ($-8.69$~eV) is about 50\% larger than 
InAs ($-5.95$~eV) \cite{cakan16}. This results in an overall increase in the HOMO-LUMO energy seen in Fig.~2(b) 
as the indium content is lowered toward GaAs. The fact that the structures are embedded into wider gap 
matrices allows us to keep track of the full variation in the $g$ factor without losing confinement up 
to energies as high as 4~eV. The opposite behavior of $g^*$ and $E_g$ will be a recurring theme also
in the following results.

\begin{figure}
\centering
\includegraphics[width=1.0\columnwidth]{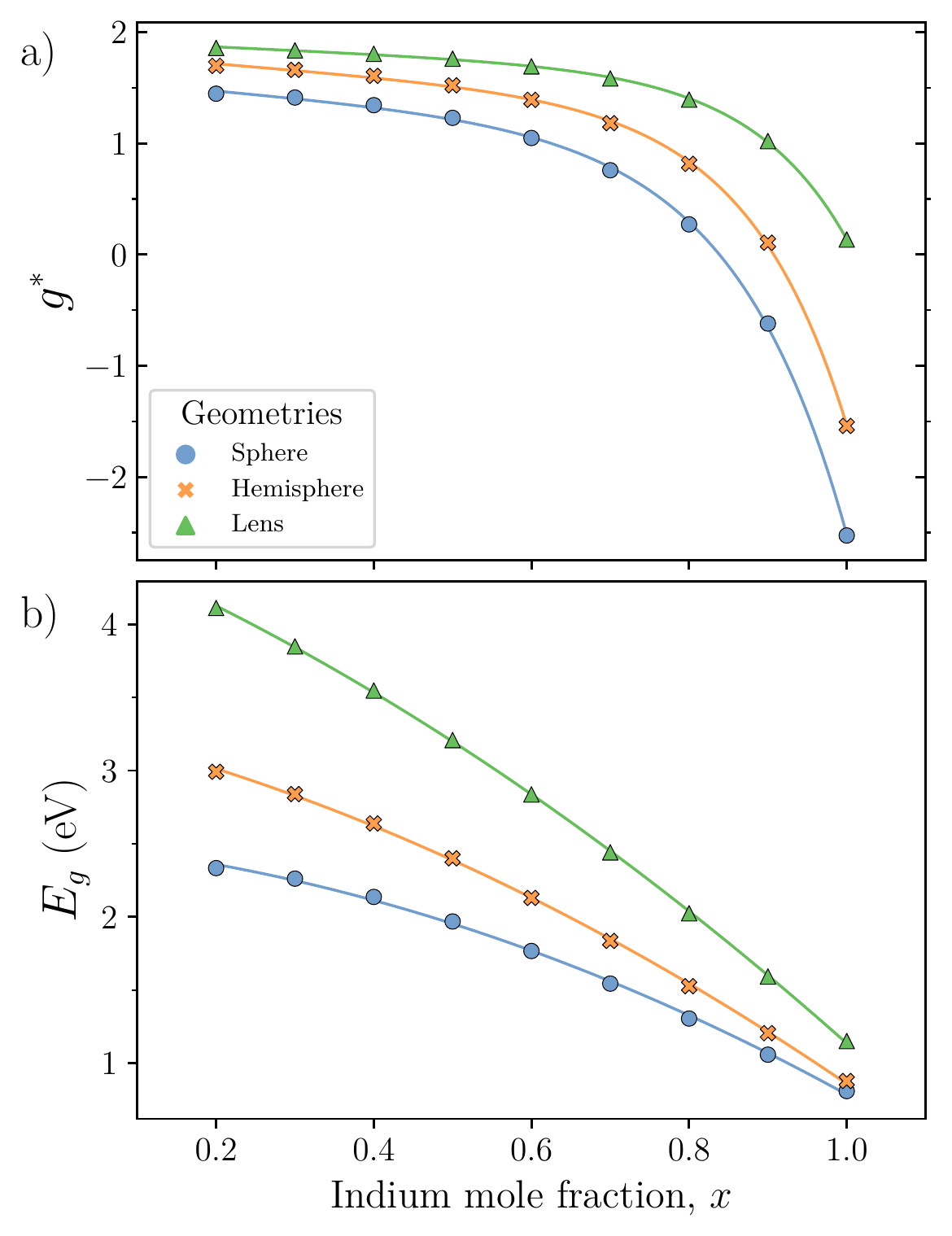}
\caption{\label{fig2} Variation of (a) $g$ factor, and (b) HOMO-LUMO energy gap, $E_g$ of the 
spherical-, hemispherical- and lens-shaped In$_x$Ga$_{1-x}$As QDs under $-2\%$ homogeneous 
strain as a function of the indium molar fraction. All of them have the same diameter of 46~nm, 
and the lens QDs have a height about 11-12~nm. Solid lines are to guide the eye.}
\end{figure}

\subsection{Dimensional dependence in lens QDs}
In the remaining sections we concentrate on the lens QDs as being the prevalent embedded self-assembled  
QD shape. First, we present in Fig.~3 the set of curves for a wide range of indium mole fractions, and 
for two different strains, all at a fixed aspect ratio (height over diameter), $h/D=0.2$.
Most notably, in Fig.~3(a) compressive strain raises the $g$ factor. The reason is predominantly 
the widening in HOMO-LUMO energy gap with compressive strain as shown in Fig.~3(b), due to negative 
band gap deformation potential of both InAs and GaAs \cite{cakan16}. Its connection with the $g$ factor 
is directly visible from the energy denominators in Eqs.~(\ref{gn}) and (\ref{roth}),
where their increase causes reduced renormalization with respect to the free-electron value.
This is reminiscent of the $k\cdot p$ conduction band effective mass expression, where $m^*$ 
decreases as the band gap increases \cite{yu-cardona}. 
A further remark is that, as in Fig.~2, for low indium concentration, $g^*$ approaches free-electron 
value and becomes largely QD diameter independent. As the indium content increases, so does the 
contribution of spin-orbit interaction, which together with decreasing energy gap both lower 
$g^*$, and instate its size dependence.

\begin{figure}
\centering
\includegraphics[width=1.0\columnwidth]{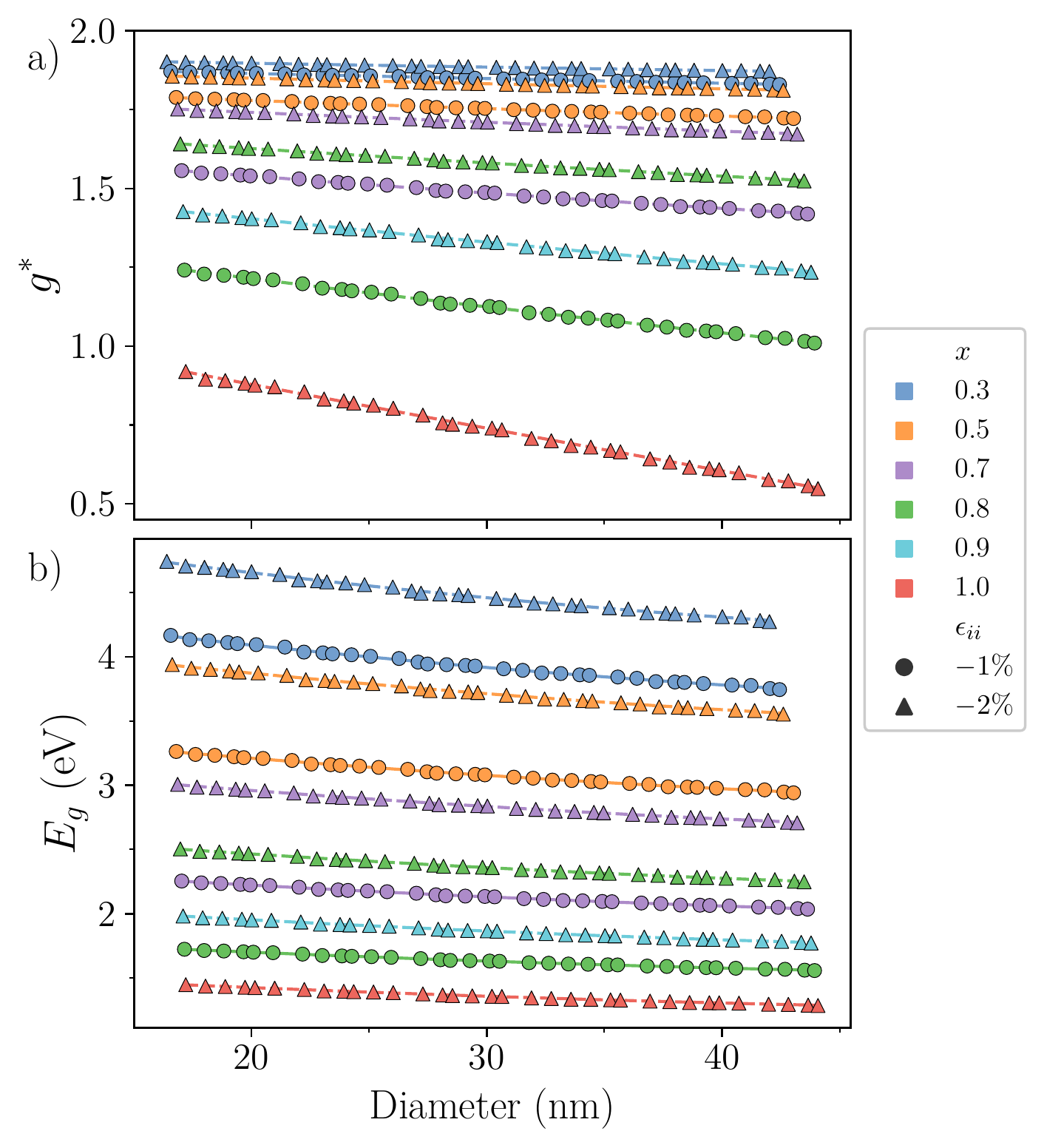}
\caption{\label{fig3} Variation of (a) $g$ factor, and (b) HOMO-LUMO energy gap, $E_g$ 
versus the diameter of the lens QD. The family of curves are all at a fixed aspect ratio of $h/D=0.2$ 
for different indium mole fraction, and strain values. Dashed lines are to guide the eye.}
\end{figure}

Another set of curves follows, this time varying the QD height, keeping the lens basal 
diameter fixed at 35~nm as shown in Fig.~4. The general trends are similar, as the 
increase in $g^*$ under compressive strain in Fig.~4(a) can be linked to that of the HOMO-LUMO energy 
gap in Fig.~4(b). In comparison to Fig.~3, there is a wider change under height, and in turn aspect ratio. 
For the considered lens diameter, $g^*\sim 0$ ensues very close to InAs composition. 
An intriguing observation is that different mole fraction and strain 
curves can perfectly overlap as in $(x=0.5, ~\epsilon_{ii} =-0.01)$ and $(x=0.8, ~\epsilon_{ii} 
=-0.03)$. This suggests that, as far as $g$ factor is concerned, there can be a universal dependence 
under a decisive parameter, as we discuss next.

\begin{figure}
\centering
\includegraphics[width=1.0\columnwidth]{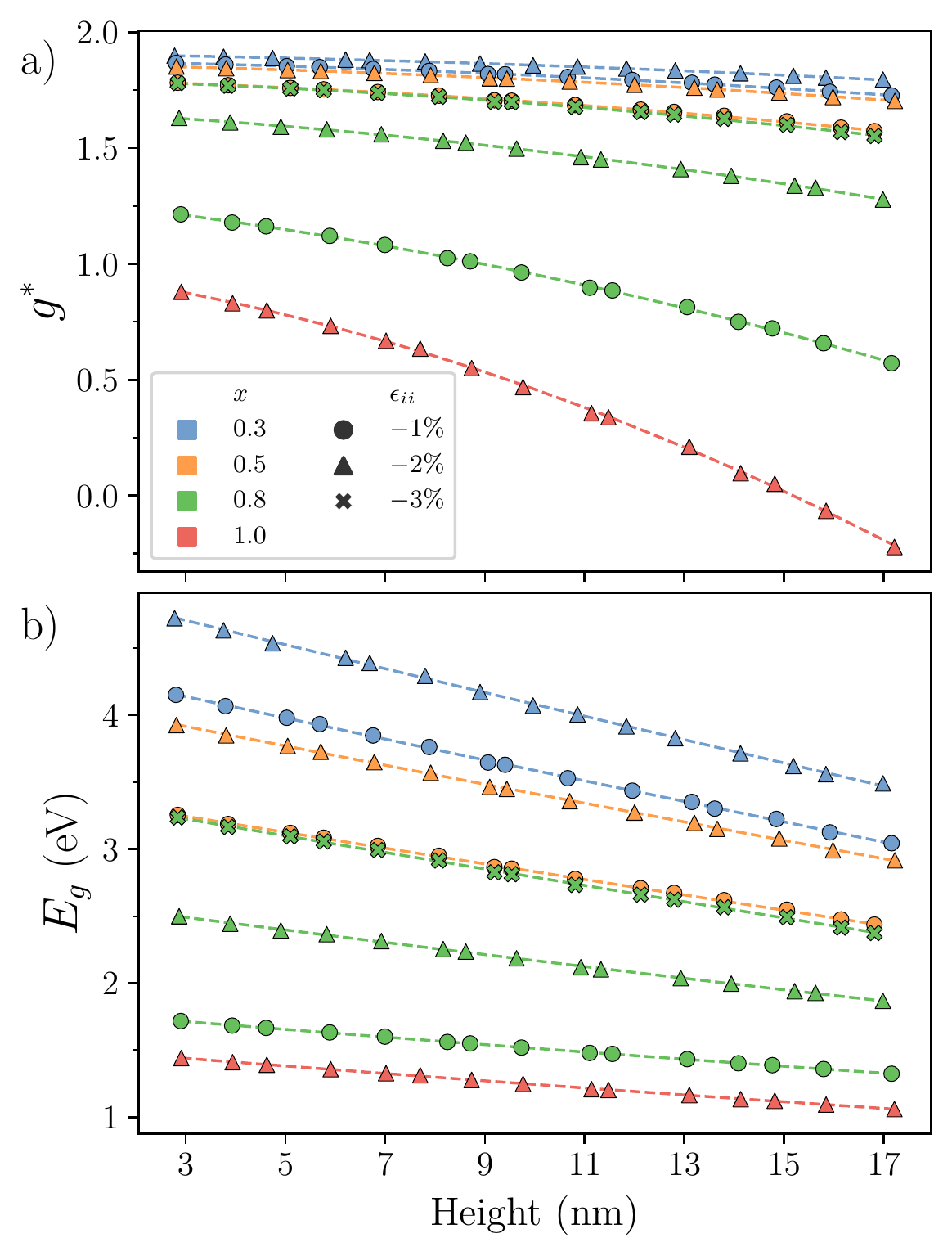}
\caption{\label{fig4} Variation of (a) $g$ factor, and (b) HOMO-LUMO energy gap, $E_g$ 
versus the height of the lens QD for a fixed basal diameter of 35~nm. The family of 
curves are for different indium mole fraction, and strain values. Dashed lines are to guide the eye.}
\end{figure}

\subsection{Universality with respect to gap energy}
We now recast all the various QD $g$-factor data above as a function of the HOMO-LUMO
energy gap, $E_g$ for each case. In connection to low-temperature magnetoluminescence 
experiments it can be easily extended to include the excitonic binding energy \cite{bulutay10}. 
When we replot the data in 
Figs.~2, 3 and 4 in this manner we observe that all the family of curves for distinct mole 
fractions, strains, aspect ratios and heights coalesce to a universal curve as presented in 
Figs.~5(a), 5(b), and 5(c), respectively. This not only supports the earlier finding of Ref.~\cite{tadjine17}, 
but also extends it to diverse geometries and alloys while allowing 
for penetration into surrounding matrix material. All these data can be well represented
by a simple curve of the same bulk form \cite{roth59,pryor06}
\begin{equation}
\label{gfit}
g^*(E_g)= 2-\frac{2.06}{E_{g}(E_{g}-0.22)}\, ,
\end{equation}
where $E_g$ is in eV. 
According to Eq.~(\ref{gfit}), we can predict that the electrons in InGaAs QDs possessing $s$-shell transition energies 
close to 1.13~eV will be least susceptible to magnetic field, due to $g^*\sim 0$.

This analysis can also be harnessed to infer some unknown values in the experimental data. To illustrate 
this point, in Fig.~5 (a) the pink star symbol corresponds to InGaAs lens QD having a diameter 
around 30~nm and height of $7-8$~nm \cite{schwan11a}, and the pink cross to another QD measured by photocurrent 
spectroscopy \cite{wu20}, for both of which only the magnitude of $g^*$ could be reported.
Using Eq.~(\ref{gfit}) we can resolve either one to be positive.
From a more general angle, this universality warrants a recipe by merely tuning the gap energy 
through any means for the pursuit of $g$-factor engineering \cite{medeiros03,nakaoka05}.

\begin{figure}
\centering
\includegraphics[width=1.0\columnwidth]{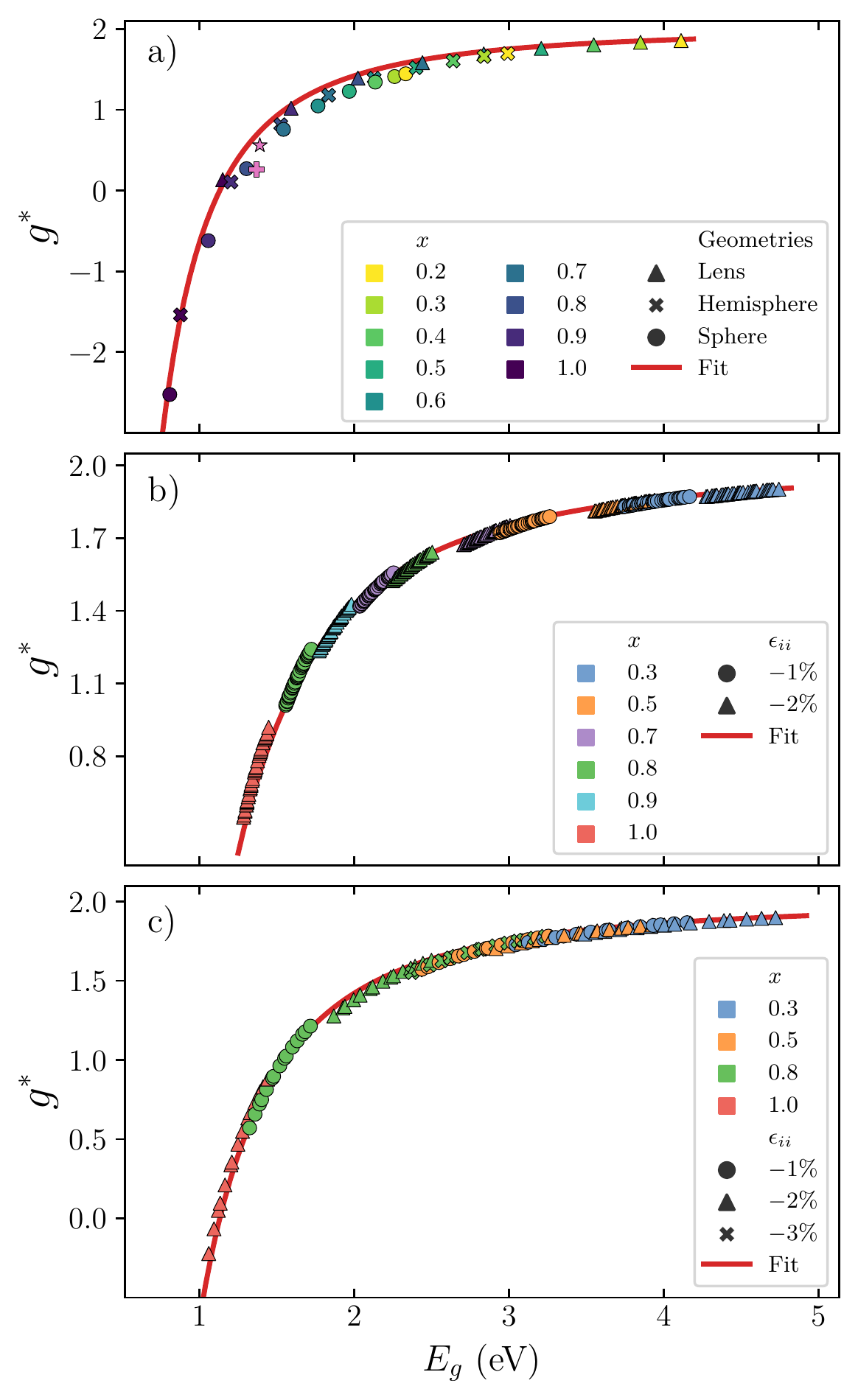}
\caption{\label{fig5} Universal $g$-factor behavior obtained when the data for
(a) different geometries in Fig.~2, (b) $D$ series in Fig.~3, and (c) $h$ series at $D=35$~nm in Fig.~4, 
are redrawn with respect to $E_g$. See the text for two other literature data points included in (a), 
as star \cite{schwan11a} and cross \cite{wu20} symbols in pink. All fitted curves obey Eq.~(\ref{gfit}).} 
\end{figure}

\section{Conclusions}
Using an empirical pseudopotential atomistic electronic structure theory, $g$-tensors of a large number 
of embedded InGaAs QDs with different shape, size, indium fraction and strain 
combinations are computed. This analysis provides the general traits of $g$-factor variation. 
For specific applications, when taken into account in their growth control or post-selection, it can 
be beneficial for achieving $g$-near-zero InGaAs QDs, or direct ESR-based quantum 
logic operations. Our study also validates a recent report based on tight-binding electronic structure 
for compound QDs that the $g$ factor acquires a universal behavior with respect to the gap energy of 
the QD regardless of its structural details \cite{tadjine17}. We additionally exhibit that this applies to 
alloy InGaAs QDs of various shapes and finite confinement allowing for penetration to the matrix. 
It remains to be examined whether these conclusions will be affected by an inhomogeneous atomic scale  
strain distribution within the QD.

\begin{acknowledgments}
This work was funded by T\"urkiye Bilimsel ve Teknolojik Ara\c{s}tirma Kurumu 
(TUBITAK) under Project No.~116F075.
The numerical calculations reported in this paper were partially performed at T\"UB\.ITAK ULAKB\.IM, 
High Performance and Grid Computing Center (TRUBA resources).
\end{acknowledgments}

\end{document}